\documentclass[11pt,a4paper]{article}
\usepackage[hyperref]{rocling2025}
\usepackage{times}
\usepackage{latexsym}

\usepackage{microtype}

\usepackage{amsmath, amssymb, amsfonts, bm}
\usepackage{cite}
\usepackage{algorithmic}
\usepackage{graphicx}
\usepackage{textcomp}
\usepackage{xcolor}
\usepackage{booktabs}
\usepackage{tabularx}
\usepackage{adjustbox}
\usepackage{multirow}
\usepackage{printlen}
\usepackage{colortbl}

\roclingfinalcopy 
\def\roclingpaperid{20} 

\title{Memory-Efficient Training for Text-Dependent SV with Independent Pre-trained Models}

\author{
  Seyed Ali Farokh \\
  Department of Computer Engineering \\
  Amirkabir University of Technology \\
  Tehran, Iran \\
  \texttt{alifarokh@aut.ac.ir} \\\And
  Hossein Zeinali \\
  Department of Computer Engineering \\
  Amirkabir University of Technology \\
  Tehran, Iran \\
  \texttt{hzeinali@aut.ac.ir} \\
}

\date{}

\begin{document}
\maketitle

\begin{abstract}
This paper presents our submission to the Iranian division of the Text-Dependent Speaker Verification Challenge (TdSV) 2024. Conventional TdSV approaches typically jointly model speaker and linguistic features, requiring unsegmented inputs during training and incurring high computational costs. Additionally, these methods often fine-tune large-scale pre-trained speaker embedding models on the target domain dataset, which may compromise the pre-trained models' original ability to capture speaker-specific characteristics. To overcome these limitations, we employ a TdSV system that utilizes two pre-trained models independently and demonstrate that, by leveraging pre-trained models with targeted domain adaptation, competitive results can be achieved while avoiding the substantial computational costs associated with joint fine-tuning on unsegmented inputs in conventional approaches. Our best system reached a MinDCF of 0.0358 on the evaluation subset and secured first place in the challenge.
\end{abstract}

\begin{keywords}
Text-dependent Speaker Verification, Speaker Verification, Memory-efficient Training, Pre-trained Models, Transfer Learning
\end{keywords}

\section{Introduction}
Speaker verification (SV) is the task of confirming an individual’s identity based on their voice. It involves comparing one or more enrollment utterances with a test utterance and can be performed in either a text-independent (TiSV) or text-dependent (TdSV) setting. In TiSV, the phonetic content of the utterances is unrestricted, and only the speaker’s identity is verified, whereas in TdSV, the system verifies both the speaker’s identity and the specific phrase spoken. With the development of various neural network architectures~\citep{xie2019utterance, desplanques2020ecapa, zeinali2019but, snyder2018tstp}, loss functions~\citep{xiang2019margin, zhang2017triplet, wang2018amsoftmax, deng2019arcface}, and pooling methods~\citep{snyder2018tstp, india2019mha, zhu2018mhsa}, TiSV has seen considerable improvement in recent years, whereas TdSV has remained relatively underexplored. TdSV systems can be either phrase-dependent (i.e., shared passphrases), where a fixed set of phrases is predefined by the system, or phrase-independent (i.e., user-defined passphrases), allowing users to customize their phrases~\citep{zeinali2024tdsv}. With the growing demand for voice-based authentication, TdSV has gained increasing attention, as the phonetic content can be used as passphrases~\citep{tu2022survey}, adding an extra layer of security to voice-based access control systems.

This paper presents our system submitted to Task 1 of the Text-dependent Speaker Verification Challenge 2024\footnote{Challenge website: https://tdsvc.github.io}~\citep{zeinali2025tdsv}, which aimed to encourage participants to explore novel approaches for TdSV. The challenge was organized into two divisions: an international one, which included two subtasks focusing on shared and user-defined passphrases, and an Iranian division, which mirrored Task 1 of the worldwide challenge but specifically emphasized developing methods with limited GPU resources. In this challenge, model enrollment is done using three enrollment utterances, and each trial consists of a test utterance and a model identifier. Speaker verification trials fall into one of the following categories:

\begin{itemize}
    \item \textbf{Target Correct (TC)}: The speaker matches the claimed model and utters the correct phrase.
    \item \textbf{Target Wrong  (TW)}: The speaker matches the claimed model but utters an incorrect phrase.
    \item \textbf{Impostor Correct (IC)}: The speaker does not match the claimed model but utters the correct phrase.
    \item \textbf{Impostor Wrong (IW)}: The speaker does not match the claimed model and utters an incorrect phrase. This category was excluded from the current year's challenge, as it does not pose sufficient difficulty for contemporary models.
\end{itemize}

In the context of TdSV, proposed systems are required to integrate both speaker and phrase verification scores and accept only TC trials\footnote{For Text-independent Speaker Verification (TiSV), the task definition differs: both TC and TW trials are accepted.}. Task 1 is phrase-dependent, employing a fixed set of ten phrases (five in Persian and five in English) for enrollment and testing. Additionally, to enhance the complexity of the challenge, some test utterances in TW trials were sourced from free-text recordings.

The primary evaluation metric adopted by TdSV 2024 is the normalized minimum Detection Cost Function (MinDCF), as defined in NIST SRE 2008 as a weighted sum of miss and false error probabilities, with $P_{target}=0.01$, $C_{FalseAlarm}=1$, and $C_{Miss}=10$. The Equal Error Rate (EER) will also be reported as a secondary performance measure.

Previous successful approaches to TdSV typically jointly model speaker characteristics and the linguistic content of utterances. For instance, \citet{liu2021phonemeaware} proposed a phoneme-aware attentive pooling method that incorporates frame-level phoneme posteriors into attentive pooling, improving the model's ability to utilize phonetic information effectively. Also, some studies have employed supervised multi-task learning to jointly learn speaker and linguistic features for further improvement~\citep{yang2020factorizedembeddings, han2021sjtu}.

However, joint speaker and phrase modeling has some drawbacks compared to independent modeling. First, model development becomes more complex than developing the system based on independent phrase and speaker embedding models. Additionally, since phrase modeling requires attending to an entire utterance, inputs cannot be chunked during training, requiring variable-length inputs to be zero-padded. This issue substantially increases GPU memory requirements, particularly for recent transformer-based models, due to their quadratic time and memory complexity~\citep{vaswani2017transformer}. 

Furthermore, as demonstrated in this work, pre-trained speaker embedding models are highly effective at extracting speaker-related features while disregarding other information in input utterances. However, when subjected to multi-task fine-tuning, these models are prone to lose their initial ability to extract speaker-related features, allocating capacity to learning linguistic content instead. This shift reduces their effectiveness, especially when in-domain data for multi-task fine-tuning is limited.

Motivated by these challenges, we leverage the full capacity of pre-trained models and develop a TdSV system based on independent pre-trained models for phrase and speaker verification. For phrase verification, we fine-tune a pre-trained cross-lingual speech representation model for bilingual automatic speech recognition (ASR) in Persian and English, followed by a further fine-tuning stage for phrase classification. This classifier is used to reject incorrect phrases. Similarly, we develop several speaker embedding extractors based on pre-trained ResNets and Whisper~\citep{radford2023whisper} for our speaker verification system. After rejecting incorrect phrases using the phrase classifier, final verification scores are obtained by computing cosine similarity between test and enrollment embeddings.

Experimental results demonstrate that with well-designed fine-tuning stages, our TdSV system built on independently pre-trained models can achieve performance comparable to systems that jointly model speaker-related and linguistic information while using only a single Nvidia RTX 3090 GPU. This strategy substantially lowers GPU memory requirements and, consequently, reduces computational costs compared to the multi-GPU setups typically employed for training speaker recognition models~\citep{zheng2023unisound}. Our best system secured first place in the Iranian division of the challenge and outperformed the third-place team in the international division~\citep{zeinali2025tdsv}.

The rest of the paper is organized as follows: Section~\ref{sec:datasets} introduces the datasets used in this work. Sections~\ref{sec:phrase_sys} and~\ref{sec:speaker_sys} describe the architecture of our phrase and speaker verification systems, respectively. The experimental results and discussion are given in Section~\ref{sec:results}, and we conclude in Section~\ref{sec:conclusion}.

\section{Challenge Datasets}
\label{sec:datasets}

The DeepMine dataset~\citep{zeinali2018deepmine, zeinali2019deepmine} is the primary source of the training and evaluation data for TdSV 2024. It was collected through crowd-sourcing, and while all participants were native Persian speakers, most contributed to the English portion of the dataset as well. The official TdSV 2024 data for Task 1 includes three subsets: training, development, and evaluation. The training subset consists of 183,431 utterances from 1,620 speakers. Among the utterances, 31,738 are free-text, while the rest were drawn from a fixed set of ten phrases comprising five Persian and five English phrases. The development and evaluation subsets are intended solely for system evaluation and contain 117,348 and 6,464,241 trials, respectively. During evaluation, model enrollment is conducted using three recordings of a specific phrase, and each trial includes a test utterance and a model identifier. The development set is provided to participants for evaluation and parameter tuning before submitting results to the official leaderboard. The evaluation subset is used for the official evaluation of the challenge. In addition to the DeepMine dataset, participants are also allowed to use the following datasets:

\begin{itemize}
    \item \textbf{VoxCeleb 1\&2}~\citep{nagrani2017voxceleb1, chung2018voxceleb2} are two large-scale datasets collected from YouTube videos, which contain over one million recordings from 7,205 celebrities. In this work, due to resource constraints, only VoxCeleb 1 was used, which includes over 100,000 utterances from 1,251 speakers.
    \item \textbf{LibriSpeech}~\citep{panayotov2015librispeech} is a standard ASR corpus in US English that comprises approximately 1,000 hours of speech from 2,338 speakers. We only used the \textit{train-clean-100} subset of this dataset to train our phrase verification system, which contains about 100 hours of speech.
    \item \textbf{Common Voice}~\citep{ardila2020cv} is a multilingual speech dataset created from contributions of volunteers from worldwide. For this challenge, teams are restricted to using the Persian (Farsi) subset, which contains approximately 363 hours of validated speech from 4,148 speakers\footnote{Common Voice 18.0, released on 6/19/2024}. To prepare this subset for training our speaker verification systems, we excluded speakers with fewer than 30 recordings. From the remaining speakers with more than 650 recordings, we randomly selected 650 utterances per speaker, resulting in a final dataset with 125,017 utterances from 813 speakers.
\end{itemize}

The challenge rules prohibit the use of any other public or private data for training.

\subsection{Data Augmentation}
\label{sec:augment}

We did not use any augmentation methods in our phrase verification system. However, following the previous successful studies on speaker verification~\citep{chen2022sjtu, zheng2023unisound}, we adopted SoX-based speed perturbation by factors of 0.9 and 1.1 to triple the number of speakers during training, followed by an on-the-fly implementation of the following augmentations, each applied with a probability of 0.6: noise addition using the MUSAN dataset~\citep{snyder2015musan}, reverberation using RIRs dataset~\citep{ko2017rirs}, and gain augmentation.
\section{Phrase Verification System}
\label{sec:phrase_sys}

Our proposed system for TdSV 2024 consists of two independent subsystems for phrase and speaker verification. The phrase verification system is a classifier that rejects TW trials, while the speaker verification system is responsible for producing similarity scores. Although this system design does not benefit from joint modeling of speaker and text, it greatly simplifies the system development process and allows for the use of various pre-trained models for each subsystem with minimal modifications.

The phrase classifier is an 11-class model trained with standard softmax. The first ten classes correspond to the set of phrases in the challenge, and the final class represents free text (or ``none of the above''). This classifier is built on XLSR\footnote{Facebook/wav2vec2-xls-r-300m}~\citep{conneau2021xlsr}, a pre-trained cross-lingual speech representation model trained by solving a self-supervised contrastive task, proven to be effective in low-resource languages compared to traditional feature extraction methods. This model takes a raw waveform as input and produces a sequence of features.

Moreover, to improve the model's ability to extract linguistic features from Persian and English inputs, we initially fine-tuned the XLSR for bilingual speech recognition in Persian and English. During this phase, 30\% of the training subset of Common Voice Farsi and LibriSpeech (\textit{train-clean-100}) were used, and the model was trained using CTC loss~\citep{graves2006ctc} for 40 epochs, with an initial learning rate of 0.001 and an effective batch size of 32. In our experiments, this phase contributes to improving the performance of the phrase verification system.

Finally, to train the classifier, an attention-based pooling layer was added to the fine-tuned XLSR to compute fixed-dimensional utterance-level feature vectors from frame-level representations $h_t$ ($t=1,...,T$):
\begin{equation}
    e_t = W_1 h_t + b_1 ,
\end{equation}
\begin{equation}
    \alpha_t = \frac{\exp(e_{t})}{\sum_{\tau}^{T}\exp(e_{\tau})} ,
\end{equation}
\begin{equation}
    \Tilde{h} = \sum_{t}^{T} \alpha_t (W_2 h_t + b_2) ,
\end{equation}
where, $e_t$ and $\alpha_t$ are the attention score and weight, respectively. $\Tilde{h}$ refers to the utterance-level feature vector, which is finally fed to a fully connected layer with ReLU activation, followed by a linear classifier. The network was trained using the Cross-Entropy loss function for one epoch on the entire training samples of the challenge dataset, with a learning rate of 0.0005 and an effective batch size of 64.
\begin{table}[t]
\centering
\renewcommand{\arraystretch}{1.0}
\begin{adjustbox}{width=\columnwidth}
\begin{tabular}{lcccccc}
    \toprule
    \toprule
    \multirow{2}{*}{\textbf{System}} & \multicolumn{3}{c}{\textbf{Full Training}} & \multicolumn{3}{c}{\textbf{Domain Adaptation}} \\
    \cmidrule(lr){2-4}
    \cmidrule(lr){5 -7}
    & Epoch & BS & LR & Epoch & BS & LR  \\
    \midrule 
    \textbf{S2}                     & -     & -     & -     & 15    & 32    & 3e-4 \\
    \textbf{S3}                     & 100   & 64    & 1e-3  & 15    & 32    & 3e-4 \\
    \textbf{S4}                     & 15    & 64    & 1e-3  & 7     & 28    & 5e-5 \\
    \textbf{S5}                     & 15    & 64    & 1e-3  & 7     & 28    & 5e-5 \\
    \bottomrule
    \bottomrule
\end{tabular}
\end{adjustbox}
\caption{Hyper-parameters used in different submitted systems S2–S5 (BS = batch size, LR = learning rate).}
\label{tab:hyperparams}
\end{table}
\begin{table*}[t]
\centering
\renewcommand{\arraystretch}{1.0}
\setlength{\tabcolsep}{10pt}
\begin{adjustbox}{width=\textwidth}
\begin{tabular}{llccccc}
    \toprule
    \toprule
    \multirow{2}{*}{\textbf{System}} & \multirow{2}{*}{\textbf{Architecture}} & \multirow{2}{*}{\textbf{Training Stages}} & \multicolumn{2}{c}{\textbf{Development}} & \multicolumn{2}{c}{\textbf{Evaluation}} \\
    \cmidrule(lr){4-5}
    \cmidrule(lr){6-7}
    & \multicolumn{2}{c}{}  & MinDCF$_{0.01}$ & EER(\%) & MinDCF$_{0.01}$ & EER(\%) \\
    \midrule 
    \textbf{S1}                     & ResNet34              &                   & 0.0614            & 1.3938                & 0.0784            & 1.7390 \\
    \textbf{S2}                     & ResNet293             & T$_2$             & 0.0225            & 0.8733                & \textbf{0.0376}   & \textbf{1.1080} \\
    \textbf{S3}                     & ResNet152             & T$_1$ + T$_2$     & 0.0191            & 0.6757                & 0.0764            & 2.3444 \\
    \textbf{S4}                     & Whisper-PMFA          & T$_1$ + T$_2$     & 0.0163            & \textbf{0.6121}       & 0.0584            & 2.0410 \\
    \textbf{S5}                     & Whisper-PMFA          & T$_1$ + T$_2$     & \textbf{0.0161}   & 0.6126                & 0.0583            & 2.0445 \\
    \midrule
    \multicolumn{3}{l}{\textbf{Fusion (S1$\sim$S5)}}     & \textbf{0.0119}   & \textbf{0.5605}   & \textbf{0.0358}       & 1.2457 \\
    \bottomrule
    \bottomrule
\end{tabular}
\end{adjustbox}
\caption{Results of different submissions on the development and evaluation sets.}
\label{tab:results}
\end{table*}
\begin{table}
\renewcommand{\arraystretch}{1.0}
\setlength{\tabcolsep}{10pt}
\centering
\begin{adjustbox}{width=\columnwidth}
\begin{tabular}{ccc}
    \toprule
    \toprule
    \textbf{Subset} & \textbf{MinDCF$_{0.01}$} & \textbf{EER(\%)} \\ 
    \midrule
    Development & 0.0000 & 0.00 \\
    Evaluation  & 0.0003 & 0.01 \\
    \bottomrule
    \bottomrule
\end{tabular}
\end{adjustbox}
\caption{Phrase verification performance on TC-vs-TW trials.}
\label{tab:phrase_results}
\end{table}
\section{Speaker Verification System}
\label{sec:speaker_sys}
To leverage the full power of pre-trained SV models and mitigate the computational cost of training randomly initialized models, we explored two directions for developing our SV system. In the first approach, we fine-tuned several pre-trained ResNet-based models, widely used as a standard architecture in speaker verification. In the second approach, we studied the performance of pre-trained ASR models adapted for SV, which have shown promising results in previous studies~\citep{zhang2022mfaconformer, cai2023pretrainingconformer, liao2023unifiedconformer}. More specifically, we employed the Whisper-PMFA~\citep{zhao2024whisperpmfa} method, which involves fine-tuning a pre-trained Whisper model for speaker recognition.

\subsection{Training Protocol}
We trained our models in two stages:

\begin{itemize}
    \item \textbf{Full training ($T_1$)}: In this stage, models were trained on a combination of out-of-domain data (Common Voice Farsi and VoxCeleb 1) and in-domain (DeepMine) data, totaling 3,684 speakers, to learn robust and generalizable speaker embeddings across different domains. Pre-trained ResNets did not undergo this stage, as they are already capable of extracting rich speaker-specific features. During this phase, 300 consecutive frames of each input utterance were randomly selected in each epoch to prevent overfitting, reduce GPU memory usage, and accelerate training. Moreover, all augmentation methods explained in Section~\ref{sec:augment} were applied. We employed the widely used AAM-Softmax~\citep{deng2019arcface} loss with the subcenter method and the Inter-TopK penalty~\citep{zhao2021speakin} to train our models, with a constant margin and scale of 0.2 and 32, respectively.
    \item \textbf{Domain adaptation ($T_2$)}: We fine-tuned our models using in-domain data after full training to bridge the domain gap and improve performance. During this stage, augmentation methods and the Inter-TopK penalty were removed to prevent domain mismatch. Additionally, the number of randomly selected frames was increased from 300 to 600 to enhance the models' generalization capability~\citep{garciaromero2019length, garciaromero2020magneto}. Fine-tuning was performed with smaller learning rates to preserve the models' generalization abilities.
\end{itemize}

All models were optimized using SGD with a momentum of 0.9 and a weight decay of 1e-4. We also utilized an exponential decay scheduler with a minimum learning rate of 5e-5 for $T_1$ and 1e-6 for $T_2$. Other training hyper-parameters are listed in Table~\ref{tab:hyperparams}. Note that gradient accumulation was used to achieve the target effective batch size when GPU memory was limited. The dimensionality of speaker embeddings was set to 256 across all models. All experiments were conducted on a single Nvidia RTX 3090 GPU using the WeSpeaker toolkit~\citep{wang2024wespeaker}.

\subsection{ResNet}
ResNet~\citep{xie2019utterance} is a widely used architecture for speaker recognition that has performed excellently in previous speaker verification challenges~\citep{zheng2023unisound}. Consequently, many open-source implementations and pre-trained models have been publicly released based on this architecture. Trained on large-scale datasets like VoxCeleb 1\&2, these pre-trained models can provide a robust starting point for training speaker recognition models on other datasets by improving their generalization and speeding up the convergence.

During the challenge period, we submitted three systems based on a bottleneck-block ResNet, all adopting temporal statistics pooling~\citep{snyder2018tstp} for aggregating variable-length sequence features into utterance-level embeddings. The first system (S1) was a pre-trained ResNet34 without domain adaptation, while the second one (S2) was a pre-trained ResNet293 that underwent domain adaptation. Finally, we applied both training stages to a randomly initialized ResNet152 to obtain our last ResNet-based system (S3).

\subsection{Whisper-PMFA}
Building on the successful use of pre-trained ASR models in speaker verification~\citep{zhang2022mfaconformer, cai2023pretrainingconformer}, \citet{zhao2024whisperpmfa} recently proposed Whisper-PMFA (Partial Multi-Scale Feature Aggregation using Whisper) to leverage the capabilities of Whisper, a large-scale multilingual ASR model based on transformer architecture. Whisper-PMFA adapts Whisper for speaker verification by selectively concatenating frame-level outputs from specific transformer layers rather than aggregating features from all layers. This approach not only reduces computational overhead but also enhances performance by minimizing the integration of irrelevant information from lower-impact layers.

Inspired by this, we studied the performance of Whisper-PMFA in this challenge. Since Whisper was not trained for the speaker recognition task, we applied both training stages to Whisper-PMFA. Additionally, before the full training stage, we froze the Whisper parameters and fine-tuned the model for five epochs to prevent updating the pre-trained model in the wrong direction due to the random initialization of newly added components. We submitted two Whisper-PMFA-based systems (S4 and S5) to this challenge, differing only in the AAM-Softmax margin used during the domain adaptation phase: 0.35 for S4 and 0.2 for S5.

\subsection{Feature Extraction}
80-dimensional log Mel filter bank energies with a 25ms window and 10ms frame-shift were extracted for our ResNet-based models. Voice activity detection (VAD) was not applied, and all features were mean-normalized. Likewise, 80-dimensional log magnitude Mel spectrograms consistent with the pre-trained Whisper were utilized for training Whisper-PMFA.

\subsection{Backend}
Speaker embeddings were extracted from the final fully connected layer of the models, and cosine similarity was used to compute scores. Since model enrollment is done using three utterances in this challenge, we used the average of embedding vectors of each model during scoring.

Afterward, AS-Norm~\citep{wang2020asnorm} was used for score normalization, using 1,620 cohorts obtained from speaker-wise averaging of all embeddings in the training subset of the challenge dataset. The top 300 most similar scores were selected to compute the mean and standard deviation for normalization.

Finally, we adopted score fusion by averaging single-system scores to further improve performance.
\begin{table}
\centering
\renewcommand{\arraystretch}{1.0}
\setlength{\tabcolsep}{8pt}
\begin{adjustbox}{width=\columnwidth}
\begin{tabular}{lccc}
    \toprule
    \toprule
    \multicolumn{1}{c}{\multirow{2}{*}{\textbf{Methods}}}   & \multicolumn{2}{c}{\textbf{Development}} \\
    \cmidrule(lr){2-3}
    \multicolumn{1}{c}{}    & MinDCF$_{0.01}$   & EER(\%) \\
    \midrule
    Whisper-PMFA (T$_1$)            & 0.0234    & 0.9253 \\
    \midrule
    + Domain adaptation (T$_2$)     & 0.0177    & 0.6273 \\
    ++ AS-Norm                      & 0.0161    & 0.6126 \\
    \bottomrule
    \bottomrule
\end{tabular}
\end{adjustbox}
\caption{Ablation study on Whisper-PMFA.}
\label{tab:ablation}
\end{table}
\section{Results}
\label{sec:results}

Table~\ref{tab:results} shows the evaluation results of our single and fusion systems on the development and evaluation subsets of the challenge after applying AS-Norm and rejecting TW trials.
The results indicate that the Whisper-PMFA method outperforms the widely used ResNet architecture with random initialization, conforming to the findings of previous studies on the effectiveness of adapting pre-trained ASR models for speaker verification. However, it can be observed from the results that the ResNets pre-trained on approximately twice the data (VoxCeleb 1\&2) can considerably surpass Whisper-PMFA after a well-designed domain adaptation stage, which highlights the importance of large-scale pre-training in improving the generalization ability of speaker verification models.

\begin{figure}[]
  \centering
  \includegraphics[width=\columnwidth]{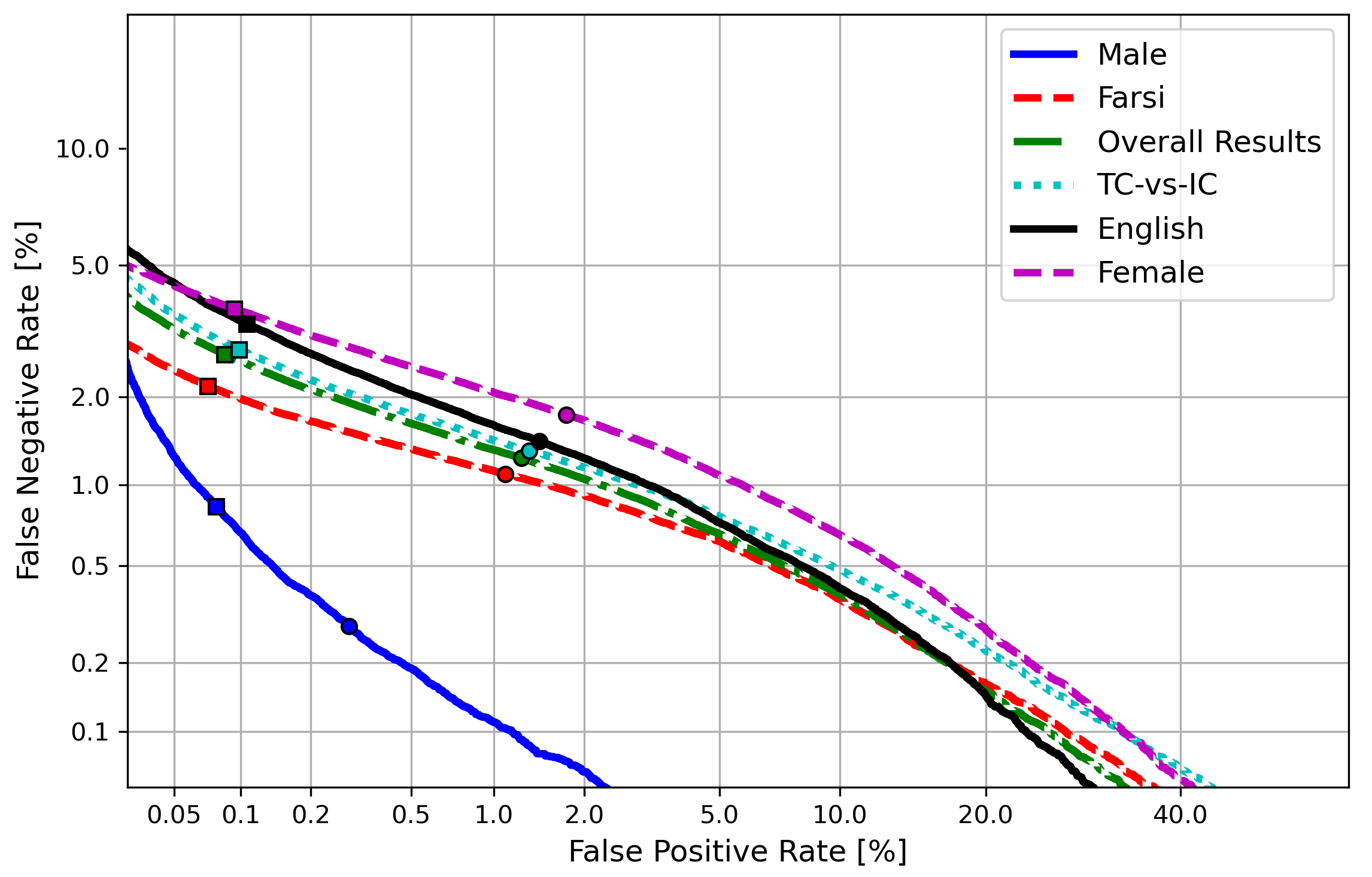}
  \caption{DET curves of our best-performing system.}
  \label{fig:det}
\end{figure}

In addition, Figure~\ref{fig:det} presents the Detection Error Tradeoff (DET) curves of the best-performing system for different categories of evaluation data. The results indicate that the model generally performs better on Persian phrases, which is expected given that the DeepMine dataset was collected from native Persian speakers, many of whom are likely less fluent in English. Furthermore, the results show noticeably higher performance for male speakers compared to female speakers. This disparity is not solely due to the inherent challenges of verifying female voices, but is also influenced by the specific characteristics of the DeepMine dataset, as discussed in its original description~\citep{zeinali2018deepmine, zeinali2019deepmine} and in the official challenge results paper~\citep{zeinali2025tdsv}.

We also report the MinDCF and EER of the proposed phrase verification system on TC-vs-TW trials of the development and evaluation subsets (Table~\ref{tab:phrase_results}). According to the results, our phrase verification system demonstrates a near-optimal performance on this task.

\subsection{Ablation Study}

We conducted an ablation study on our Whisper-PMFA system (S5). The development set of the challenge dataset was used as our evaluation benchmark. We can observe from the results (Table~\ref{tab:ablation}) that the domain adaptation phase improved the MinDCF from 0.0234 to 0.0177. Also, a further improvement of MinDCF to 0.0161 was achieved after applying AS-Norm.

\subsection{Comparison with Other Teams}

To contextualize our performance, we report in Table~\ref{tab:teams} the evaluation results of our best system alongside the top-performing submissions in Task 1 of the international division of the TdSV Challenge. Team names and scores are taken directly from the official challenge results paper~\citep{zeinali2025tdsv}, which also provides brief descriptions and comparisons of the proposed architectures. As shown, our system achieves a lower MinDCF than the team ranked third in the international division.
\vspace{-1.5em}
\begin{table}
\centering
\renewcommand{\arraystretch}{1.0}
\setlength{\tabcolsep}{10pt}
\begin{adjustbox}{width=\columnwidth}
\begin{tabular}{lcc}
    \toprule
    \toprule
    \textbf{Team}                   & \textbf{MinDCF$_{0.01}$}   & \textbf{EER(\%)} \\ 
    \midrule
    Team 04~\citep{sreekanth2024slt} & \textbf{0.0297}   & 1.132 \\
    Team 08                  & 0.0326            & \textbf{1.013} \\
    \textbf{Our System}      & 0.0358            & 1.246 \\
    Team 02                  & 0.0379            & 1.164 \\
    Team 01                  & 0.0504            & 2.245 \\
    \bottomrule
    \bottomrule
\end{tabular}
\end{adjustbox}
\caption{Evaluation results for our best system and the top-ranked teams in Task 1 of the international division of TdSV.}
\label{tab:teams}
\end{table}
\section{Conclusion}
\label{sec:conclusion}

In this paper, we present our system for Task 1 of the Iranian division of the Text-dependent Speaker Verification (TdSV) Challenge 2024, focusing on resource-constrained training for TdSV systems. Unlike previous methods that jointly model speaker-related and linguistic features, our approach leverages two independent pre-trained models for phrase and speaker verification. This design reduces the computational costs associated with joint modeling during training while fully utilizing the capabilities of pre-trained models to achieve competitive performance. Our best system achieved a MinDCF of 0.0358 on the evaluation subset, securing first place in the challenge.

\vspace{-1.5em}
\bibliography{rocling2025}
\bibliographystyle{acl_natbib}

\end{document}